\documentstyle[aps,multicol,graphics,epsfig]{revtex}

\newcommand{\ket}[1]{\left| #1 \right\rangle}

\begin{document}
\draft

\title{Universal Quantum  Cloning in Cavity QED}
\author{P. Milman$^{1}$\thanks{Electronic address: Perola.Milman@lkb.ens.fr},
        H. Ollivier$^{2,3}$\thanks{Electronic address: harold.ollivier@polytechnique.org}, and J. M. Raimond$^1$}
\address{$^1$ Laboratoire Kastler Brossel,
            D\'epartement de Physique de l'Ecole Normale Sup\'erieure, \\
            24 rue Lhomond, F-75231 Paris Cedex 05 France}
\address{$^2$ LANL, T-6 MS B288, Los Alamos NM 87544, USA.}
\address{$^3$ INRIA-Projet CODES, B.P. 105, F-78153 Le Chesnay Cedex, France.}
\address{$^4$ Coll\`ege de France, 11 place Marcelin-Berthelot, F-75005,
            Paris, France}
\maketitle

\begin{abstract}
We propose an implementation of an universal quantum
cloning machine [UQCM, Hillery and Buzek, Phys. Rev. A {\bf 56}, 3446 (1997)]
in a Cavity Quantum Electrodynamics (CQED) experiment. This UQCM 
acts on the electronic states of atoms that interact with the
electromagnetic field of a high $Q$ cavity.
We discuss here the specific case of the $1 \rightarrow 2$ cloning process
using either a one- or a two-cavity configuration.
\end{abstract}

\begin{multicols}{2}
\paragraph{Introduction.}
Quantum theory provides new and unexpected effects when
compared to classical physics. Among them, the no-cloning theorem, derived in
1982 by Wooters and Zurek\cite{WZ_1982a}, plays a particularly important role: 
While classical information can be copied perfectly and many times,
quantum information cannot. This fundamental difference is a
consequence of the unavoidable creation of quantum correlations.
Since perfect cloning is not possible, an important question naturally
arises : What is the best quantum copying operation? The answer to
this question is context-dependent. On the one hand, there is a single
transformation that produces the best identical copies of a qubit prepared in any input states.
This ``universal quantum cloning machine" (UQCM) has been discussed for the first 
time in\cite{Buzek:PRA1996}. On the
other hand, many other rules of the game can be considered, such as state
dependent cloning\cite{Bruss:PRA1998}, cloning of
3-dimensional states\cite{Gisin:quantph} and cloning of
orthogonal qubits\cite{Sofyan:quantph}.

The  quality of a copy is usually measured by the quantum
fidelity \cite{Jozsa:JMO1994}. This quantity is discussed,
in the context of universal quantum cloning machine (UQCM),
in \cite{Buzek:PRA1996} and \cite{Gisin:PRL1997}. When $M$ copies are produced 
from $N$ identical pure
2-dimensional states, the fidelity of the copies is given by $F(N,M)
= (NM+N+M)/(M(N+2))$. For the simplest case
of two copies produced from one input state, this expression reduces to $F(1,2)=5/6$. 
The complete understanding of the fidelity 
behavior versus $N$ and $M$ is still a subject of
debate, with connections to the measurement
and state estimation problems~\cite{Banaszek:PRL2001}.
Beyond these fundamental problems, the interest of
quantum cloning machines also encompasses a wide area of quantum
information processing, including quantum cryptography, teleportation \cite{Grosshans:PRA2001}, 
eavesdropping, state preservation and measurement-related
problems, as well as quantum algorithm improvements
\cite{Ernesto:PRA2000}.

The derivation of the optimal UQCM transformation led to  several
proposals \cite{Simon:PRL2000} for its experimental implementation. 
Most of them, based on the Buzek
and Hillery quantum logics network \cite{Buzek:PRA1997}, use the
quantum optics framework. 
Experimental quantum cloning has been realized up to now only with photons
as the carriers of quantum information. This information was either encoded
in different degrees of freedom of the same photon (polarization and position)  
\cite{Huang:PRA2001} or in the photon
polarization only \cite{Antia:Science2002}.
An alternative network adapted to NMR-based quantum information processors has
also been proposed and experimentally implemented \cite{Cummins:PRL2002}.

In this paper, we propose an implementation of the $1 \rightarrow 2$ 
UQCM operating for atomic states  in the Cavity QED (CQED) context \cite{Raimond:RMP2001}.
The quantum information is coded on electronic levels of long-lived highly
excited Rubidium (Rb) atoms. Our protocol realizes, with four atoms, the transformation described in
\cite{Buzek:PRA1996}, with an original quantum logics network based on the resonant
interaction between the atoms and two high-$Q$ niobium superconducting microwave cavities $C_a$ and $C_b$.
We discuss, at the end of this paper, an adaptation of the scheme using two different modes of a single 
cavity \cite{Rauschenbeutel:PRA2001}, making the proposal implementation more realistic with the present
cavity QED set-up. This paper focuses on the quantum logics protocol. The interested reader
can find more details about the experimental techniques in \cite{Raimond:RMP2001}

Let us first recall the optimal $1 \rightarrow 2$ UQCM transformation \cite{Buzek:PRA1996}.
When the qubits are encoded in the basis $\{\ket{+}, \ket{-}\}$, the UQCM performs the transformations
\begin{eqnarray}
\label{machine}
\ket{-}\ket{{\cal B}} & \rightarrow &\sqrt{2 \over 3}\ket{-}\ket{-}\ket{{\cal A}}+\sqrt{1 \over 3}\ket \Phi \ket{{\cal A}_{\perp}}  \nonumber \\
\ket{+}\ket{{\cal B}} & \rightarrow &\sqrt{2 \over 3}\ket{+}\ket{+}\ket{{\cal A}_{\perp}}+\sqrt{1 \over 3}\ket \Phi \ket{{\cal A}},
\end{eqnarray}
where the first ket of the {\it l.h.s} represents the input qubit and 
$\ket{{\cal B}}$ is the initial state of the blank copies and of possible ancilla qubits involved in the process.
In the {\it r.h.s}, the first two kets are the quantum clones, 
$\ket \Phi = (\ket + \ket - + \ket - \ket +)/\sqrt 2$. The third ket represents two possible orthogonal final states,
 $\ket{{\cal A}}$ and $\ket{{\cal A}_{\perp}}$ for the ancilla qubits.

Our scheme makes use of three atomic levels, $\ket e$, $\ket g$ and $\ket i$.
The transition between levels $\ket{e}$ and $\ket{g }$ can be set in and out of resonance 
with the cavity modes, using the Stark effect
induced by an electric field applied between the Fabry Perot cavity mirrors \cite{Raimond:RMP2001}.
The auxiliary level $\ket{i}$ is far off-resonant
from the cavity fields and is not coupled to them. However,
it can be accessed via classical microwave pulses either from level $\ket g$ (one photon transition) or from level $\ket e$ (two-photon transition). 
The atomic qubit encoding is
$\ket{+}=1/\sqrt{2}(\ket{i}+\ket{g})$ and
$\ket{-}=1/\sqrt{2}(\ket{i}-\ket{g})$. The photon number states of 
each cavity mode is denoted as  $\ket{n}_i$, where $i=(a,b)$.

\paragraph{Description of the protocol:}

\begin{figure}[h]
\center
\includegraphics[width=3.2in, height=2in]{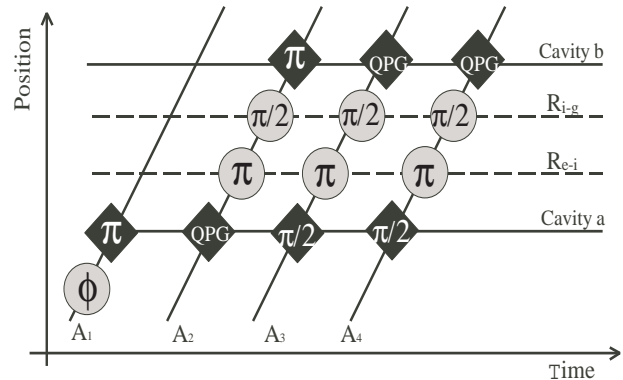}
\caption{Detailed scheme of the atom field interactions in each cavity. $A_1$
enters the first cavity in state $\sqrt{1/3}\ket{e}_1+\sqrt{2/3}\ket{g}_1$ via a classical pulse of a certain duration and
transfers its state to the cavity field in a $\pi$  Rabi pulse. The cavity performs
thus a QPG in $A_2$, the atom carrying the state to be cloned
$\alpha\ket{+}_2+\beta\ket{-}_2$. After being manipulated by microwave classical
fields, $A_2$ also transfers its state to the second cavity, which is now
entangled to the first one. A third atom $A_3$, prepared in state $\ket{g}_3$,
crosses the first cavity performing a $\pi/2$ Rabi pulse. The first cavity field's
state is completely recovered by the atomic states via the passage of a fourth
atom, $A_4$, also prepared in $\ket{g}_4$, which makes a $\pi$ Rabi pulse. Both of
these atoms will interact resonantly with the second cavity field after
classical microwave pulses manipulation. They will perform a resonant QPG, which
will leave the total combined atom+second cavity field in the desired final
state, corresponding to the cloning transformation. } \label{fig1}
\end{figure}

The sequence of operation achieving the UCQM transformation is pictorially depicted
in Fig. (\ref{fig1}). It presents, in a space-time diagram, the space lines of the two
cavity modes and of the four Rydberg atoms, $A_{1-4}$, involved in the process. The atom-cavity 
resonant interactions are represented by black lozenges. Classical microwave pulses mixing the
atomic levels are represented as gray circles.

The cavity fields are initially prepared in the vacuum state $\ket{0}_i$ \cite{Raimond:RMP2001}. 
The first atom, $A_1$, intially in state $\ket{g}_1$, is prepared in state
\begin{equation}\label{a1}
\ket{\Psi}_1=\sqrt{2 \over 3}\ket{g}_1+\sqrt{1 \over 3}\ket{e}_1,
\end{equation}
by a classical pulse resonant with the $\ket g\rightarrow\ket e$ transition. The coefficients in the equation above are set adjusting the duration of the classical pulse to a $\phi = \arcsin(\sqrt{1/3})$ rotation (see Fig. 1).  
The production of (\ref{a1}) can be checked in auxilliary experiments
measuring the population of states $\ket{g}_1$ and $\ket{e}_1$ and the quantum coherence. 
The atomic state (\ref{a1}) is then transferred to $C_a$,
through a $\pi$ pulse of resonant quantum Rabi oscillation \cite{Maitre:PRL1997}. Atom
$A_1$ finally leaves $C_a$ in state $\ket{g}_1$ and the cavity field is left in 
state  $\ket{\psi}_a=\sqrt{2/3}\ket{0}_a+\sqrt{1/3}\ket{1}_a$.
The first atom's final state being factored out, it will no longer be considered here.

Atom $A_2$, carrying the state to be cloned, crosses then $C_a$. 
It is prepared in the arbitrary state
\begin{equation}\label{a2}
\ket{\Psi}_2=\alpha\ket{+}_2+\beta\ket{-}_2,
\end{equation}
where $\alpha$ and $\beta$ are complex coefficients.
Note that the preparation of this state, which is not part of the quantum 
cloning process, is not represented in figure 1. This atom
interacts with the cavity field, performing a $2\pi$ quantum
Rabi pulse, amounting to a resonant quantum
phase gate (QPG) described in \cite{Rauschenbeutel:PRL1999}.
The QPG produces a $\pi$ phase shift of the atom-cavity quantum state 
if and only if the atom is in state $\ket{g}$ and
the cavity in state $ {\ket 1}_a$. 
When expressed in the $\left \{ \ket{+}, \ket{-}\right \}$ basis, this QPG
operation amounts to a controlled not gate (CNOT),
where the control qubit is the field state. 
After this interaction the total entangled atom-field state becomes 
\begin{equation}\label{cnot1}
\sqrt{2 \over 3}(\alpha\ket{+}_2+\beta\ket{-}_2)\ket{0}_a+\sqrt{1 \over
3}(\alpha\ket{-}_2+\beta\ket{+}_2)\ket{1}_a\ .
\end{equation}

We then send a third atom, $A_3$, prepared in state $\ket{g}_3$. It
interacts resonantly with $C_a$, for a
time interval corresponding to a $\pi/2$ quantum Rabi pulse, producing the
state
\begin{eqnarray}\label{a3}
&&\sqrt{2 \over 3}(\alpha\ket{+}_2+\beta\ket{-}_2)\ket{g}_3\ket{0}_a\\
&&+\sqrt{1 \over
6}(\alpha\ket{-}_2+\beta\ket{+}_2)(\ket{g}_3\ket{1}_a+\ket{e}_3\ket{0}_a).
\nonumber
\end{eqnarray}
The state of $C_a$ is finally transferred to a fourth atom,
$A_4$, initially in $\ket{g}_4$ via a resonant $\pi$ quantum Rabi pulse, 
creating the three-atom entangled state:
\begin{eqnarray}\label{a4}
&&\sqrt{2 \over 3}(\alpha\ket{+}_2+\beta\ket{-}_2)\ket{g}_3\ket{g}_4\\
&&+\sqrt{1 \over
6}(\alpha\ket{-}_2+\beta\ket{+}_2)(\ket{g}_3\ket{e}_4+\ket{e}_3\ket{g}_4),
\nonumber
\end{eqnarray}
and leaving $C_a$ in the vacuum state, which factors out. 

Classical microwave pulses then address the three atoms, transforming $\ket{e}$
into $\ket{i}$ via a two-photon $\pi$ pulse. This transformation does not affect state $\ket g$.
Then, another classical $\pi/2$ pulse is applied on the three atoms,
combining states $\ket{g}$ and $\ket{i}$. The sequence of
transformations produced by these classical pulses can be summarized as follows:
\begin{eqnarray}\label{class}
&&\ket{g}\longleftrightarrow\ket{g}\longleftrightarrow\sqrt{1
\over 2}\left (\ket{i}-\ket{g} \right )=\ket{-}, \nonumber \\
&&\ket{e}\longleftrightarrow\ket{i}\longleftrightarrow\sqrt{1 
\over 2}\left (\ket{i}+\ket{g} \right )=\ket{+}.
\end{eqnarray}

The remaining part of the protocol involves the second cavity $C_b$.
Atom $A_2$ interacts resonantly
with $C_b$ for a time interval
corresponding to a $\pi$ quantum Rabi pulse, transferring its state to the field mode. 
The final state of $A_2$ is $\ket g_2$ and also factorizes out. The
total state of $A_3$,  $A_4$ and $C_b$ is then :
\begin{eqnarray}\label{cav21}
&&\sqrt{2 \over 3}\left (\alpha\ket{1}_b+\beta\ket{0}_b \right
)\ket{-}_3\ket{-}_4\\
&&+\sqrt{1 \over 6}\left (\alpha\ket{0}_b+\beta\ket{1}_b \right )\left
(\ket{-}_3\ket{+}_4+\ket{+}_3\ket{-}_4 \right ) \nonumber
.
\end{eqnarray}
Atoms $A_3$ and $A_4$ interact then independently and succesively with $C_b$.
They  perform a resonant QPG, corresponding to a
CNOT in the $\left \{ \ket{+},\ket{-} \right \}$  basis. The final state of
these three systems writes, after a simple rearrangement of terms:
\begin{eqnarray}\label{finalb}
&&\alpha \left [ \sqrt{2 \over 3}\ket{+}_3\ket{+}_4\ket{{\cal
A}}+\sqrt{1 \over 3}{\ket \Phi}_{3,4} \ket{{\cal A}_{\perp}} \right] \\
&+&\beta \left [ \sqrt{2 \over 3}\ket{-}_3\ket{-}_4\ket{{\cal
A}_{\perp}}+\sqrt{1 \over 3}{\ket \Phi}_{3,4} \ket{{\cal A}}\right] \nonumber,
\end{eqnarray}
where $\ket{{\cal A}}=\ket{g}_1\ket{g}_2\ket{0}_a\ket{0}_b$, $\ket{{\cal
A}_{\perp}}=\ket{g}_1\ket{g}_2\ket{0}_a\ket{1}_b$ and ${\ket \Phi}_{3,4} = 
(\ket +_3 \ket -_4 + \ket -_3 \ket +_4)/\sqrt 2$.
In Eq.~(\ref{finalb}), the second cavity field ensures the orthogonality of 
$\ket {\cal A}$ and $\ket {\cal A_\perp}$ and hence is the important qubit in the ancilla's final state.
This achieves the implementation of the optimal $1 \rightarrow 2$ cloning process.

Eq. (\ref{finalb}) shows that this sequence actually implements the UCQM transformation 
given by Eq.~(\ref{machine}). In this proposal, the blank state 
$\ket{{\cal B}}$ corresponds to the initial
state of  atoms $A_1$, $A_3$ and $A_4$ and of both cavity fields. It writes thus
\begin{equation}\label{b}
\ket{{\cal B}}=(\sqrt{1 \over 3}\ket{e}_1+\sqrt{2 \over
3}\ket{g}_1)\ket{g}_3\ket{g}_4\ket{0}_a\ket{0}_b.
\end{equation}

\paragraph{Discussion:}

We can now discuss the feasibility of an experimental
implementation of the UQCM in a CQED system. The basic operations
(quantum gates and classical field pulses) involved in the scheme 
have already been troughfully tested \cite{Raimond:RMP2001}. Their 
implementation thus does not present any major difficulty. The availability of an
experimental configuration with two cavities can be considered 
as natural development of the present configurations using a single cavity, and mainly a matter of time. Note also some other
interesting proposals require at least a two-cavity system \cite{Luiz:PRA1985}.

Atoms interact with $C_a$ or $C_b$ for a time corresponding at most to a 
$2 \pi$ quantum Rabi pulse. The single photon Rabi frequency \cite{Maitre:PRL1997} 
being $\Omega/2\pi=50$~kHz, the atomic velocity should be
 $\approx 500$~m/s, in the range used in present experiments. Cavity and atomic
relaxation are of course important issues. The circular Rydberg atoms lifetime is much
longer than the protocol duration and is not bound to be a limiting factor. The main
cause of decoherence in the present set-up is the cavity mode relaxation.  
The quantum information is stored in $C_a$ only during the time interval
between the passage of $A_1$ and $A_4$. Each atom may enter the cavity immediately after the
preceding one has left it. The total quantum information storage time is of the order of
4 full atomic transit times, i.e. $\approx 2.10^{-4}$~s. This is shorter than present
cavity damping times (about 1 ms). The cavity $C_b$ stores quantum information for
an even shorter time interval. Note finally that the atomic transit time between the two cavities
does not matter to evaluate the influence of damping, since the quantum information is then
carried by long-lived atomic systems.

An alternative implementation of our UCQM scheme uses two modes of a single cavity. In the present
experimental set-up, the cavity sustains two gaussian modes, $M_a$ and $M_b$, with orthogonal linear polarizations.
Due to mirrors imperfections, these two modes have slightly different resonant frequencies 
(splitting 130 kHz). Since this splitting is much larger than the atom-field coupling $\Omega$, the
atoms resonantly interact with one mode only at at same time. Stark tuning can be used to tailor atomic interactions
with the two modes during the atomic transit time through the cavity.

In this scheme, $A_1$ leaves its state in $M_a$. Then, $A_2$
performs the CNOT operation in $M_a$. It is set off-resonance with both modes
for a short time interval during which the microwave classical pulses
are applied. Atom $A_2$ it then tuned to resonance with $M_b$ for its final quantum Rabi pulse.
$A_3$ and $A_4$ interact first resonantly with $M_a$, undergo the classical pulses while
being off-resonance from the two modes and finally interact with $M_b$ as described above.
This implementation of the UCQM, requiring a single cavity, would be much simpler to
realize. Each atom should interact with the cavity for a total time corresponding at most to
a $3 \pi$ quantum Rabi pulse (the duration of the classical pulses is negligible). The atomic
velocity should be about 330 m/s, still well within the available range. The quantum information
is stored in the cavity modes for a slightly longer time than in the two-cavity arrangement (four
times the full transit time of atoms at the slower 330 m/s velocity). Cavity damping should thus
be somewhat smaller.

The UQCM operation verification can, in principle, be performed by the usual detection
techniques \cite{Raimond:RMP2001}. As mentioned above, the fidelity is,
ideally,$5/6$ while the trivial
production of a maximally mixed state gives an average  fidelity of $2/3$. This means
that the fidelity should be measured with a precision 
greater than $\approx 92\% \quad(= 1 - \frac{1}{2}(5/6-2/3))$.
Note that, in the NMR quantum cloning experiment \cite{Cummins:PRL2002},
 this degree of precision was not reached, so that the improvement due to the cloning
process could not be verified. 

In our proposal, all the elementary operations, quantum Rabi or classical field
pulses, are prone to errors. The total number of these operations is sixteen if we 
take into account the detection and preparation process. The necessary precision could only be reached if each pulse
has a fidelity greater than $\sqrt[16]{0.92}$. This value,being about $0.995$ 
is still out
of the experimental reach (present pulse imperfections are between 3 and 10\%). This figure, 
however, sets an interesting goal to be reached.

\paragraph{Conclusion.}

We described a protocol implementing the universal optimal copying
transformation in CQED. Basic quantum information
operations have already been implemented in this context 
\cite{Rauschenbeutel:PRL1999}, and proposals that could
extend these experimental realizations to more elaborated quantum
information algorithms \cite{Yamaguchi:PRA2002} are naturally appealing.  

The quantum logics network used in our scheme is simpler than previous ones
by making use of auxiliary degrees of freedom which are
discarded in the end of the process. Note also that the same protocol 
can be applied to the cloning of
equatorial qubits \cite{Bruss:PRA1998}, i.e., $\ket{\psi}=\sqrt{1 /2}\left
(\ket{0}+e^{i\phi}\ket{1}\right)$ by sending $A_1$ in state $\sqrt{1/ 2}\left
(\ket{e}+\ket{g}\right)$.

The authors wish to thank J.
Kempe for calling our attention to the problem and M. Brune and S. Haroche for inspiring discussions.

Laboratoire Kastler Brossel, Universit\'{e} Pierre et Marie Curie and ENS, is
associated with CNRS (UMR 8552).


\vspace*{2cm}

\end{multicols}


\begin{thebibliography}{99}


\bibitem{WZ_1982a}
        W. K. Wooters and W. H. Zurek, Nature (London) {\bf 299}, 802(1982); D.
Dieks, Phys. Lett. {\bf 92A}, 271 (1982).
\bibitem{Buzek:PRA1996}
        V. Buzek and M. Hillery, Phys. Rev. A {\bf 54}, 1844 (1996).

\bibitem{Bruss:PRA1998}
        D. Bruss, {\it et al.}, Phys. Rev. A {\bf 57}, 2368 (1998); D. Bruss
{\it et al.} Phys. Rev. A {\bf 62}, 012302 (2000).



\bibitem{Gisin:quantph}
        N. J. Cerf, T. Durt, and N. Gisin, quant-ph/0110092 (2001).

\bibitem{Sofyan:quantph}
        J. Fiurasek, S. Ibsidir, S. Massar, and N. J. Cerf, quant-ph/0110016
(2001).
\bibitem{Jozsa:JMO1994} R. Josza, J. Mod. Optics {\bf 41}, 2315 (1994).


\bibitem{Gisin:PRL1997}
        N. Gisin and S. Massar, Phys. Rev. Lett. {\bf 79}, 2153 (1997).



\bibitem{Banaszek:PRL2001}
        K. Banaszek, Phys. Rev. Lett. {\bf 86}, 1366 (2001); D. Bruss, A. Ekert
and C. Macchiavello, Phys. Rev. Lett. {\bf 81} , 2598 (1998).

\bibitem{Grosshans:PRA2001}
        F. Grosshans and Ph. Grangier, Phys. Rev. A {\bf 64}, 010301(R) (2001).



\bibitem{Ernesto:PRA2000}
        E. F. Galv\~ao and L. Hardy, Phys. Rev. A {\bf 62}, 022301 (2000).


 \bibitem{Simon:PRL2000}
        C. Simon, G. Weihs, and A. Zeilinger, Phys. Rev. Lett. {\bf 84}, 2993
(2000);
        J. Kempe, C. Simon, and G. Weihs, Phys. Rev. A {\bf 62}, 032302 (2000).

\bibitem{Buzek:PRA1997}
        V. Buzek, S. L. Braunstein, M. Hillery, and D. Bruss, Phys. Rev. A {\bf
56}, 3446 (1997).

\bibitem{Huang:PRA2001}
        Y.-F. Huang {\it et al.}, Phys. Rev. A {\bf 64}, 012315 (2001).

\bibitem{Antia:Science2002}
        A. Lamas-Linares, C. Simon, J. Howell, and D. Bouwmeester, Science {\bf
286}, 712  (2002);S. Fasel {\it et al}, quant-ph/0203056 (2002).

\bibitem{Cummins:PRL2002} H. K. Cummins {\it et al.}, Phys. Rev. Letters {\bf
88}, 187901 (2002).


\bibitem{Raimond:RMP2001}
        J. M. Raimond, M. Brune, and S. Haroche, Rev. Mod. Phys. {\bf 73}, 565
(2001).

\bibitem{Rauschenbeutel:PRA2001}
        A. Rauschenbeutel, P. Bertet, S. Osnaghi, G. Nogues, M. Brune, J. M.
Raimond, and S. Haroche,  Phys. Rev. A {\bf 64}, 050301(R) (2001); P. Bertet, S.
Osnaghi, P. Milman, A. Auffeves, P. Maioli, M. Brune, J. M. Raimond, and S.
Haroche, Phys. Rev. Lett. {\bf 88}, 143601 (2002).


\bibitem{Maitre:PRL1997}
        X. Ma\^itre {\it et al.}, Phys. Rev. Lett. {\bf 79}, 769 (1997); M. Brune {\it et al.}, Phys. Rev. Lett. {\bf 76}, 1800 (1995). 




\bibitem{Rauschenbeutel:PRL1999}
        A. Rauschenbeutel {\it et al.}, Phys. Rev. Lett. {\bf 83}, 5166 (1999).

\bibitem{Luiz:PRA1985} L. Davidovich, N. Zagury, M. Brune, J. M. Raimond and S.
Haroche, Phys. Rev. A {\bf 50}, R895 (1994).


\bibitem{Yamaguchi:PRA2002}
        F. Yamaguchi, P. Milman, M. Brune, J. M. Raimond, and S. Haroche, to be
published in Phys. Rev. A; M. O. Scully and M. S. Zubairy, Phys. Rev. A {\bf
65}, 052324 (2002).






\end{thebibliography}
\end{document}